\documentclass{elsart}

\usepackage{epsfig}

\newcommand{\be}{\begin{equation}}
\newcommand{\ee}{\end{equation}}
\newcommand{\bea}{\begin{eqnarray}}
\newcommand{\eea}{\end{eqnarray}}

 \def\bean{\begin{eqnarray*}}
 \def\eean{\end{eqnarray*}}

 \def\l{\left}
 \def\r{\right}

 \def\bm#1{\mbox{\boldmath$#1$}}
 \def\gsim{\mathrel{\rlap{\lower0.2em\hbox{$\sim$}}\raise0.2em\hbox{$>$}}}
 \def\ksim{\mathrel{\rlap{\lower0.2em\hbox{$\sim$}}\raise0.2em\hbox{$<$}}}

\begin{document}

\begin{frontmatter}
\title{Parton-Hadron-String Dynamics at Relativistic Collider energies}
\author[unif,fias]{E. L. Bratkovskaya\corauthref{cor1}}
\ead{Elena.Bratkovskaya@th.physik.uni-frankfurt.de}
\author[unig]{W. Cassing,}
\author[unig]{V. P. Konchakovski,}
\and
\author[unif]{O. Linnyk}
\corauth[cor1]{corresponding author}
\address[unif]{Institut f\"ur Theoretische Physik, %
  JWG Universit\"{a}t Frankfurt, D--60438 Frankfurt am Main,
  Germany}
\address[fias]{Frankfurt Institut for Advanced Studies,
         Frankfurt University, 60438 Frankfurt-am-Main, Germany}
\address[unig]{Institut f\"ur Theoretische Physik, %
  Universit\"at Giessen, %
  Heinrich--Buff--Ring 16, %
  D--35392 Giessen, %
  Germany}

\begin{abstract}
The novel Parton-Hadron-String Dynamics (PHSD) transport approach is
applied to nucleus-nucleus collisions at RHIC energies with respect
to differential hadronic spectra in comparison to available data.
The PHSD approach is based on a dynamical quasiparticle model for
partons (DQPM) matched to reproduce recent lattice-QCD results
from the Wuppertal-Budapest group in
thermodynamic equilibrium.  The transition from partonic to hadronic
degrees of freedom is described by covariant transition rates for
the fusion of quark-antiquark pairs or three quarks (antiquarks),
respectively, obeying flavor current-conservation, color neutrality
as well as energy-momentum conservation.  Our dynamical studies for heavy-ion
collisions at relativistic collider energies are compared
to earlier results from the Hadron-String Dynamics (HSD) approach -
incorporating no explicit dynamical partonic phase - as well as to
experimental data from the STAR, PHENIX, BRAHMS and PHOBOS
collaborations
for Au+Au collisions at the top RHIC energy of $\sqrt{s}$ = 200 GeV.
We find  a reasonable reproduction of hadron rapidity distributions
and transverse mass spectra and also a fair description of
the elliptic flow of charged hadrons as a function of the
centrality of the reaction and the transverse momentum $p_T$.
Furthermore, an approximate quark-number scaling of the elliptic
flow $v_2$ of hadrons is observed in the PHSD results, too.

 \end{abstract}


\end{frontmatter}

\section{Introduction}
Present experiments at the Relativistic Heavy-Ion Collider (RHIC)
have reached for short time scales the conditions met in the first
micro-seconds in the evolution of the universe after the 'Big Bang'.
The 'Big Bang' scenario implies that on these time scales the entire
state has emerged from a partonic system of quarks, antiquarks and
gluons -- a quark-gluon plasma (QGP) -- to color neutral hadronic
matter consisting of interacting hadronic states (and resonances) in
which the partonic degrees of freedom are confined. The nature of
confinement and the dynamics of this phase transition  is still an
outstanding question of todays physics. Early concepts of the QGP
were guided by the idea of a weakly interacting system of massless partons
which might be described by perturbative QCD (pQCD). However,
experimental observations at RHIC indicated that the new medium
created in ultrarelativistic Au+Au collisions is interacting more
strongly than hadronic matter (cf.\ \cite{QM01} and Refs.\ therein).
It is presently widely accepted that this medium is an almost
perfect liquid of partons \cite{Shuryak,Thoma,Andre,PT,STARS,Miklos3} as extracted
experimentally from the strong radial expansion and the scaling of
the elliptic flow $v_2(p_T)$ of mesons and baryons with the number of
constituent quarks and antiquarks \cite{STARS,SCALING2,SCALING1}. In order to explore the
transport properties of
this partonic medium microscopic studies based on non-equilibrium
dynamics are mandatory.

A consistent dynamical approach for the description of strongly
interacting systems - also out-of equilibrium - can be formulated on
the basis of Kadanoff-Baym
(KB) equations \cite{KBaym,Sascha1} or off-shell transport equations
in phase-space representation, respectively
\cite{Sascha1,Juchem,Knoll1,Crev}. In the KB theory the field quanta
are described in terms of dressed propagators with complex
selfenergies. Whereas the real part of the selfenergies can be
related to mean-field potentials (of Lorentz scalar, vector or
tensor type), the imaginary parts  provide information about the
lifetime and/or reaction rates of time-like 'particles'
\cite{Andre}. Once the proper (complex) selfenergies of the degrees
of freedom are known, the time evolution of the system is fully
governed  by off-shell transport equations (as described in Refs.
\cite{Sascha1,Juchem,Knoll1,Crev}). The determination/extraction of
complex selfenergies for the partonic degrees of freedom has been
performed before in Refs. \cite{Andre,Cassing06,Cassing07} by
fitting lattice QCD (lQCD) 'data' within  the Dynamical
QuasiParticle Model (DQPM). In fact, the DQPM allows for a simple
and transparent interpretation of lattice QCD results for
thermodynamic quantities as well as correlators and leads to
effective strongly interacting partonic quasiparticles with broad
spectral functions. We stress that mean-field potentials for the
'quarks' and 'gluons' as well as effective interactions can be
extracted from lQCD within the DQPM (cf. Ref.
\cite{Cassing07}). For a review on off-shell transport theory and
results from the DQPM in comparison to lQCD we refer the reader to
Ref. \cite{Crev}.

In preceding works \cite{PRC08,NPA09} two of the authors have presented first
Parton-Hadron-String-Dynamics (PHSD) transport calculations for
expanding partonic fireballs as well as nucleus-nucleus collisions
at Super-Proton-Synchrotron (SPS) energies of 40 to 160 A GeV. The
studies in Ref. \cite{PRC08} have addressed expanding partonic
fireballs of ellipsoidal shape in coordinate space that hadronize
according to local covariant transition rates. It was found that the
resulting hadronic particle ratios turn out to be in line with those from a
grandcanonical partition function at temperature $T \approx 170$ MeV
rather independent from the initial temperature of the partonic
system. Furthermore, the scaling of elliptic flow with initial
spatial eccentricity indicated a dynamical evolution of the system
close to ideal hydrodynamics, which so far has been successfully employed
for the description of experimental data at RHIC
\cite{Heinz1,Heinz2,Heinz3,Heinz4}\footnote{For dissipative hydrodynamics
with a finite shear viscosity $\eta$ we refer the reader to Refs.
\cite{Paul1,Paul2,Paul3,Paul4,Paul5,Heinz5,Rischke1,Rischke2,Rischke3}.}.
Additionally, the application of PHSD to
nucleus-nucleus collisions at SPS energies has demonstrated a good
reproduction of a large set of data \cite{NPA09} improving the
quality of the description within the Hadron-String Dynamics (HSD)
approach  which lacks explicit partonic degrees of freedom.
These general properties of PHSD results for idealized systems are
well in line with global observations of experiments at  RHIC
energies \cite{STARS}, however, the
actual question is about the description of various experimental
differential observables in particular at top RHIC energies where a
rather precise experimental control is possible.

The paper is organized as follows: In Section 2 we briefly review
the PHSD approach, recall the input parameters and specify the extensions
relative to Ref. \cite{NPA09}. Section 3 is
devoted to actual applications for Au + Au collisions at RHIC
energies in comparison to experimental data from the STAR, PHENIX,
BRAHMS and PHOBOS collaborations. Section 4 concludes this study with a
summary, discussion of open problems and an outlook.

\section{The PHSD approach}
The Parton-Hadron-String-Dynamics (PHSD) approach is a microscopic
covariant transport model that incorporates effective partonic as
well as hadronic degrees of freedom and involves a dynamical
description of the hadronization process from partonic to hadronic
matter. Whereas the hadronic part is essentially equivalent to the
conventional HSD approach \cite{Ehehalt,HSD} the partonic dynamics
is based on the Dynamical QuasiParticle Model (DQPM)
\cite{Cassing06,Cassing07,Andre05} which describes QCD properties in
terms of single-particle Green's functions (in the sense of a
two-particle irreducible (2PI) approach). In Ref. \cite{NPA09} we
have fitted the (essentially three) DQPM parameters for the
temperature-dependent effective coupling to the lattice QCD results
of  Ref. \cite{Cheng08} which lead to a critical temperature $T_c
\approx$ 192 MeV that corresponds to a critical energy density of
$\epsilon_c \approx$ 1.25 GeV/fm$^3$. These lattice QCD results
disagree with the lQCD 'data' by the Wuppertal-Budapest group
\cite{Wupper} - as pointed out explicitly in the summary of Ref.
\cite{NPA09} - and the conflict meanwhile has come close to an end
\cite{wupper2}.

In this respect in the present study
we refitted the DQPM parameters for the strong coupling in
order to reproduce the lattice QCD results from Ref.
\cite{aori10}. The latter yield a critical temperature $T_c
\approx$ 160 MeV and a critical energy density $\epsilon \approx$
0.5 GeV/fm$^3$ which are significantly lower than those deduced from the
results of Ref. \cite{Cheng08}. Furthermore, the
scaled interaction measure $(\epsilon - 3P)/T^4$ (with $P$
denoting the pressure) is also significantly lower in the lQCD calculations
from Ref. \cite{aori10} than in the
calculations from Cheng et al. \cite{Cheng08}. As will be shown below, the readjustment
of the DQPM parameters to the lQCD data of Ref. \cite{aori10} results in a smaller
dynamical width of the quasiparticles,  whereas the dynamical pole masses become slightly
larger. For the actual procedure of fixing the effective (temperature dependent)
strong coupling $g(T/T_c)$ we refer the reader
to Section 2.1 of Ref. \cite{NPA09}.

One might worry that the quasiparticle properties - fixed in
thermal equilibrium - also should be appropriate for out-off
equilibrium configurations. This question is nontrivial and can only be answered
by detailed model investigations e.g. on the basis of Kadanoff-Baym
equations. We recall that such studies have been summarized in
Ref. \cite{Crev} for strongly interacting scalar fields that initially are
far off-equilibrium and simulate momentum distributions of colliding systems
at high relative momentum. The
results for the effective parameters $M$ and $\gamma$, which correspond to the time-dependent pole
mass and width of the propagator, indicate that the quasiparticle
properties - except for the very early off-equilibrium
configuration - are close to the equilibrium mass and width even
though the phase-space distribution of the particles is far from
equilibrium (cf. Figs. 8 to 10 in Ref. \cite{Crev}). Accordingly,
we will adopt the equilibrium quasiparticle properties also for
phase-space configurations out-off equilibrium as  appearing in
relativistic heavy-ion collisions. The reader has to keep in mind
that this approximation is far from being arbitrary, however, not
fully equivalent to the exact solution.

\subsection{Quasiparticle properties and thermodynamics within the
DQPM}
\begin{figure}[tbh]
\centerline{\includegraphics*[width=95mm]{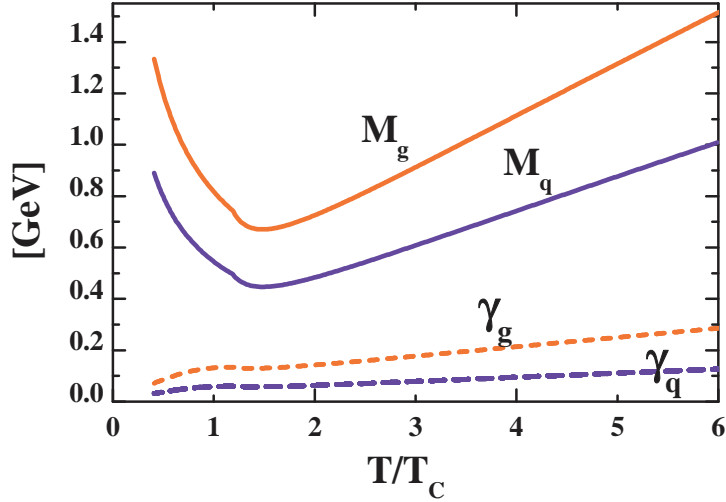}}
\caption{ The effective gluon mass
$M_g$ and witdh $\gamma_g$ as function of the scaled temperature $T/T_c$ (red lines).
The blue lines show the corresponding quantities for quarks.} \label{fig1}
\end{figure}

The actual gluon mass $M_g$ and width $\gamma_g$ -- employed as input in
the further calculations -- as well as the quark mass $M_q$ and width $\gamma_q$
are depicted in Fig. \ref{fig1}  as a
function of $T/T_c$. These values for the masses are slightly
larger than those presented in Ref. \cite{NPA09} and the width $\gamma_g$
as well as the width $\gamma_q$ are smaller due to the lower scaled interaction measure
in Ref. \cite{aori10} compared to the interaction measure in Ref. \cite{Cheng08}. This implies that the
partons become better 'quasiparticles' since the ratios
$\gamma_g/M_g$ and $\gamma_q/M_q$ decrease relative to Ref.
\cite{NPA09}. Note that for $\mu_q$ = 0 the DQPM gives
\begin{equation} \label{qma}
M_q = \frac{2}{3} M_g, \hspace{1cm} \gamma_q = \frac{4}{9}
\gamma_g \ . \end{equation}

\begin{figure}[t]
\centerline{\includegraphics*[width=92mm]{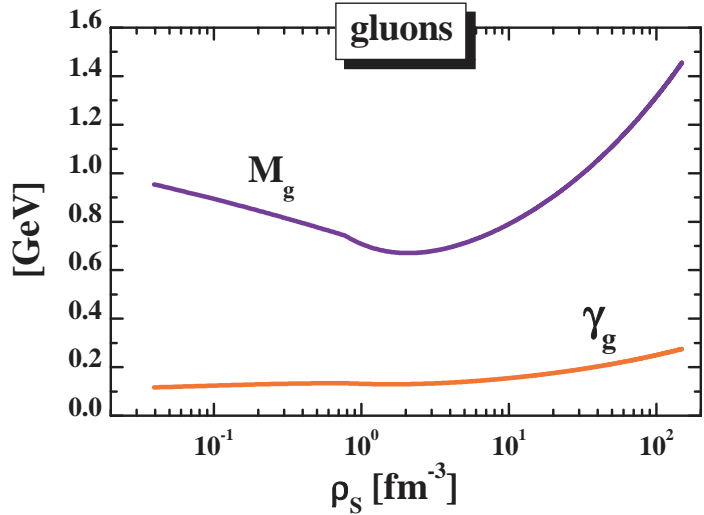}}
\caption{ The effective gluon mass
$M_g$ and width $\gamma_g$ as function of the scalar density $\rho_s$ within the
DQPM. The corresponding mass and width for quarks (for $\mu_q$ = 0)
is obtained from Eq. (\ref{qma}). Note the logarithmic scale in $\rho_s$.} \label{fig2}
\end{figure}

These variations of the masses with the temperature $T$ that appear drastic in Fig. \ref{fig1} become, however,
rather smooth if viewed as a function of the scalar parton density
$\rho_s$ defined (in thermal equilibrium) by
\begin{equation} \label{rhos}
\rho_s(\frac{T}{T_c}) = d_g \int_0^\infty  \frac{d\omega}{2 \pi}
\int \frac{d^3 p}{(2 \pi)^3}\ 2  \sqrt{p^2} \ \rho_g(\omega, {\bf p})
\ n_B(\omega/T) \ \Theta(p^2) \end{equation} $$+ d_q \int_0^\infty  \frac{d\omega}{2 \pi}
\int \frac{d^3 p}{(2 \pi)^3} \ 2  \sqrt{p^2} \  \rho_q(\omega, {\bf p})
\ n_F((\omega-\mu_q)/T) \ \Theta(p^2)$$
 $$+ d_{\bar q} \int_0^\infty  \frac{d\omega}{2 \pi}
\int \frac{d^3 p}{(2 \pi)^3} \ 2  \sqrt{p^2} \  \rho_{\bar q}(\omega, {\bf p})
\ n_F((\omega+\mu_q)/T) \ \Theta( p^2) \ ,$$
where $n_B$ and $n_F$ denote the Bose and Fermi
functions, respectively, while $\mu_q$ stands for the quark chemical potential.
The number of transverse gluonic degrees
of freedom is $d_g=16$ while the fermion degrees of freedom amount
to $d_q=d_{\bar q}=2 N_c N_f=18$ in case of three flavors
($N_f$=3). The function $\Theta( p^2)$ (with $p^2 = \omega^2 - {\bf p}^2$)
projects on time-like four-momenta since only this fraction of the four-momentum distribution
can be propagated within the light cone.
In Eq. (\ref{rhos}) the parton spectral functions $\rho_j$ are
no longer $\delta-$ functions in the invariant mass squared but
taken as \cite{Andre}
\begin{eqnarray}
 \rho_j(\omega)
 =
 \frac{\gamma_j}{ E_j} \l(
   \frac{1}{(\omega-E_j)^2+\gamma_j^2} - \frac{1}{(\omega+E_j)^2+\gamma_j^2}
 \r)
 \label{eq:rho}
\end{eqnarray} separately for quarks and gluons ($j=q,\bar{q},g$).
With the convention $E^2(\bm p) = \bm p^2+M_j^2-\gamma_j^2$, the
parameters $M_j^2$ and $\gamma_j$ are directly related to the real
and imaginary parts of the  retarded self-energy, e.g. $\Pi_j =
M_j^2-2i\gamma_j\omega$. The spectral function (\ref{eq:rho}) is
antisymmetric in $\omega$ and normalized as \begin{equation}
\label{normalize}
\int_{-\infty}^{\infty} \frac{d \omega}{2 \pi} \ \omega \
\rho_j(\omega, {\bf p}) = \int_0^{\infty} \frac{d \omega}{2 \pi} \ 2 \omega \
\rho_j(\omega, {\bf p}) = 1 \ .
\end{equation}
The dependence of the gluon mass $M_g$ and width $\gamma_g$ as a function of
$\rho_s$ (within the DQPM) is displayed in Fig. \ref{fig2} and
demonstrates that the explicit variation with $\rho_s$ is rather
moderate in view of the logarithmic scale in $\rho_s$. Note that in
transport theory the scalar forces on a 'particle' are given by
the ratio of the particle mass over its energy times the gradient of the scalar mean-field $U_s(x)$.
The latter gradient is conventionally written as $\nabla U_s(x) =
d U_s/d \rho_ s \nabla \rho_s(x)$ which demonstrates the separation
of geometry - expressed by $\nabla \rho_s(x)$ - from the strength
of the force determined by $d U_s/d \rho_ s$.

\begin{figure}[t]
\vspace{0.5cm}
\centering \includegraphics*[width=92mm]{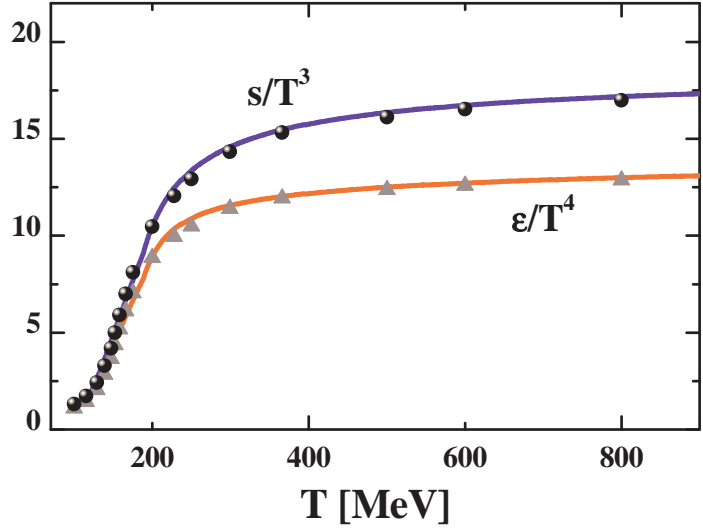} \caption{
The scaled entropy density $s(T)/T^3$ (blue line) and scaled energy
density $\epsilon(T)/T^4$ (red line) from the DQPM in comparison to
the lQCD results from Ref. \cite{aori10} (full dots and triangles).} \label{fig4}
\end{figure}

\begin{figure}[tb]
\centering \includegraphics*[width=85mm]{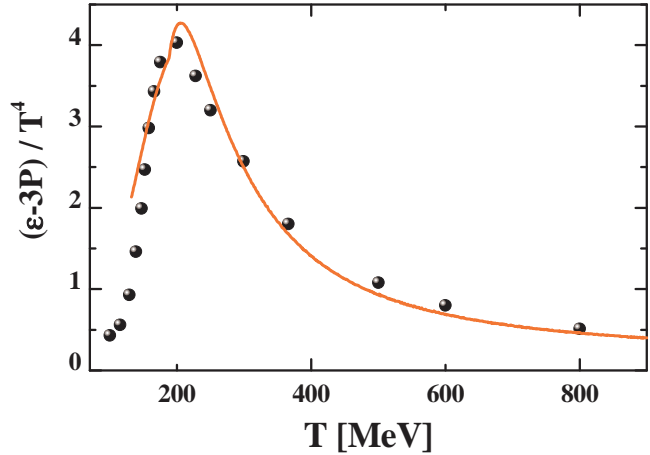} \caption{ The
dimensionless 'interaction measure' $(\epsilon - 3 P)/T^4$ within the DQPM in comparison to the
lQCD results from Ref. \cite{aori10} (full dots).} \label{fig5}
\end{figure}

With the quasiparticle properties (or propagators) fixed (cf. Fig. \ref{fig1} and Eq. (\ref{eq:rho})) one can
evaluate the entropy density $s(T)$, the pressure $P(T)$ and
energy density $\epsilon(T)$ in a straight forward manner (cf. Ref.
\cite{NPA09}). A direct comparison of the resulting entropy density $s(T)$ and energy
density $\epsilon(T)$ from the DQPM with lQCD results from
Ref. \cite{aori10} is presented in Fig. \ref{fig4}. Both results
have been divided by $T^3$ and $T^4$, respectively, to
demonstrate the scaling with temperature. We briefly note that the
agreement is sufficiently good. This also holds for the
dimensionless 'interaction measure', i.e. $(\epsilon - 3 P)/T^4$ as
demonstrated in Fig. \ref{fig5}.

We recall that the DQPM uniquely defines a potential energy density
\begin{equation} \label{Vp}
V_p(T,\mu_q) = T^{00}_{g-}(T,\mu_q) + T^{00}_{q-}(T,\mu_q) + T^{00}_{{\bar q}-}(T,\mu_q) \end{equation}
where the different contributions $T^{00}_{j-}$ correspond to the space-like part
of the energy-momentum tensor component $T^{00}_{j}$ of parton $j
= g, q, \bar{q}$ (cf. Section 3 in Ref. \cite{Cassing07}). As
demonstrated in Ref. \cite{Cassing07} this quantity is practically
independent on the quark chemical potential (for moderate $\mu_q$)
when displayed as a function of the scalar density $\rho_s$ instead of $T$ and $\mu_q$
separately. Note that the field quanta involved in (\ref{Vp}) are
virtual and thus correspond to partons exchanged in interaction
diagrams.

\begin{figure}[tb]
\centering \includegraphics*[width=85mm]{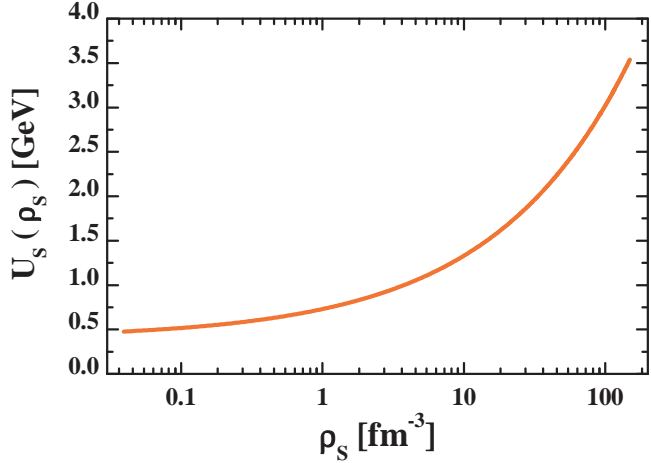} \caption{ The
scalar mean field (\ref{uss}) for quarks and antiquarks from the
DQPM as a function of the scalar parton density $\rho_s$ (2).}
\label{figpot}
\end{figure}

A scalar mean-field $U_s(\rho_s)$ for quarks and antiquarks can be defined by the
derivative,
\begin{equation} \label{uss}
U_s(\rho_s) = \frac{d V_p(\rho_s)}{d \rho_s} ,
\end{equation}
which is evaluated numerically within the DQPM. The actual
result for the new parameter-set is displayed in Fig. \ref{figpot}
as a function of the parton scalar density $\rho_s$ and shows that
the scalar mean field is in the order of a few GeV for $\rho_s >
10$ fm$^{-3}$. The mean-field (\ref{uss}) is employed in the PHSD
transport calculations and determines the force on a quasiparticle
$j$, i.e.
$ \sim M_j/E_j \nabla U_s(x) = M_j/E_j \
d U_s/d \rho_ s \ \nabla \rho_s(x)$ where the scalar density
$\rho_s(x)$ is determined numerically on a space-time grid (see
below).

We point out that in general the quasiparticle masses $M_j$ as well
as the widths $\gamma_j$ should depend also on the
four-momentum $q$ relative to the medium at rest and approach the
perturbative values at high $q^2$. So far, the
momentum-dependence of the complex self energy cannot reliably be
extracted from the lQCD results in thermodynamic equilibrium which
are essentially sensitive to momenta in the order of a few times
the temperature. This is presently an open issue and will have to
be re-addressed in future.

\subsection{Hadronization}
The hadronization, i.e. the transition from partonic to hadronic
degrees of freedom, was presented in Refs. \cite{PRC08,NPA09} and
is described below again in more detail for
clarity. The hadronization is implemented in PHSD by local covariant
transition rates as introduced first in Ref. \cite{PRC08} e.g. for
$q+\bar{q}$ fusion to a mesonic state $m$ of four-momentum $p= (\omega, {\bf
p})$ at space-time point $x=(t,{\bf x})$:
\begin{eqnarray}
&&\phantom{a}\hspace*{-5mm} \frac{d N_m(x,p)}{d^4x d^4p}= Tr_q
Tr_{\bar q} \
  \delta^4(p-p_q-p_{\bar q}) \
  \delta^4\left(\frac{x_q+x_{\bar q}}{2}-x\right) \nonumber\\
&& \times \omega_q \ \rho_{q}(p_q)
   \  \omega_{\bar q} \ \rho_{{\bar q}}(p_{\bar q})
   \ |v_{q\bar{q}}|^2 \ W_m(x_q-x_{\bar q},(p_q-p_{\bar q})/2) \nonumber \\
&& \times N_q(x_q, p_q) \
  N_{\bar q}(x_{\bar q},p_{\bar q}) \ \delta({\rm flavor},\, {\rm color}).
\label{trans}
\end{eqnarray}
In Eq. (\ref{trans}) we have introduced the shorthand notation,
\begin{equation}
Tr_j = \sum_j \int d^4x_j \int \frac{d^4p_j}{(2\pi)^4} \ ,
\end{equation}
where $\sum_j$ denotes a summation over discrete quantum numbers
(spin, flavor, color); $N_j(x,p)$ is the phase-space density of
parton $j$ at space-time position $x$ and four-momentum $p$.  In
Eq. (\ref{trans}) $\delta({\rm flavor},\, {\rm color})$ stands
symbolically for the conservation of flavor quantum numbers as
well as color neutrality of the formed hadronic state $m$ which can be
viewed as a color-dipole or 'pre-hadron'.  Furthermore, $v_{q{\bar
q}}(\rho_p)$ is the effective quark-antiquark interaction  from
the DQPM  (displayed in Fig. 10 of Ref. \cite{Cassing07}) as a
function of the local parton ($q + \bar{q} +g$) density $\rho_p$
(or energy density). Furthermore, $W_m(x,p)$ is the dimensionless phase-space
distribution of the formed 'pre-hadron', i.e.
\begin{equation} \label{Dover} W_m(\xi,p_\xi) =
\exp\left( \frac{\xi^2}{2 b^2} \right)\ \exp\left( 2 b^2 (p_\xi^2- (M_q-M_{\bar
q})^2/4) \right)
\end{equation} with $\xi = x_1-x_2 = x_q - x_{\bar q}$ and $p_\xi = (p_1-p_2)/2
= (p_q - p_{\bar q})/2$ (which had been previously introduced in
Eq. (2.14) of Ref. \cite{Dover}). The width parameter $b$ is fixed
by $\sqrt{\langle r^2 \rangle} = b$ = 0.66 fm (in the rest frame) which
corresponds to an average rms radius of mesons. We note that the
expression (\ref{Dover}) corresponds to the limit of independent
harmonic oscillator states and that the final hadron-formation
rates are approximately independent of the parameter $b$ within
reasonable variations. By construction the quantity (\ref{Dover})
is Lorentz invariant; in the limit of instantaneous 'hadron
formation', i.e. $\xi^0=0$, it provides a Gaussian dropping in the
relative distance squared $({\bf r}_1 - {\bf r}_2)^2$. The
four-momentum dependence reads explicitly (except for a factor
$1/2$)
\begin{equation} (E_1 - E_2)^2 - ({\bf p}_1 - {\bf p}_2)^2 -
(M_1-M_2)^2 \leq 0
\end{equation} and leads to a negative argument of the second
exponential in (\ref{Dover}) favoring the fusion of partons with
low relative momenta $p_q - p_{\bar q}= p_1-p_2$.

Related transition rates (to Eq. (\ref{trans})) are defined for the
fusion of three off-shell quarks ($q_1+q_2+q_3 \leftrightarrow B$)
to a color neutral baryonic ($B$ or $\bar{B}$) resonances of finite
width (or strings) fulfilling energy and momentum conservation as
well as flavor current conservation (cf. Section 2.3 in Ref.
\cite{NPA09}). In contrast to the familiar
coalescence models \cite{Koal1,Koal2} and their recent extensions
\cite{Dyn,Bleicher,Bass3,CMKo} this hadronization scheme solves the problem of simultaneously
fulfilling all conservation laws and the constraint of entropy production.

\subsection{Numerical aspects}
On the hadronic side PHSD includes explicitly the  baryon octet and
decouplet, the $0^-$- and $1^-$-meson nonets as well as selected
higher resonances as in HSD \cite{Ehehalt,HSD}. Hadrons of higher
masses ($>$ 1.5 GeV in case of baryons and $>$ 1.3 GeV in case of
mesons) are treated as 'strings' (color-dipoles) that  decay to the
known (low-mass) hadrons according to the JETSET algorithm
\cite{JETSET}. We discard an explicit recapitulation of the string
formation and
decay and refer the reader to the original work \cite{JETSET} or
Ref. \cite{Falter}.

The dynamical evolution of the system is entirely described by the
transport dynamics in PHSD incorporating the off-shell propagation
of the partonic quasiparticles according to Refs.
\cite{Sascha1,Juchem,Crev} as well as the transition to resonant
hadronic states (or 'strings') via Eq. (\ref{trans}). The time
integration for the testparticle-equations of motion (cf. Refs.
\cite{Juchem}) is performed in the same way as in case of hadronic
off-shell transport where (in view of the presently  momentum-independent width
$\gamma$) the simple relation (19) in Ref. \cite{NPA807} is
employed. For the collisions of partons two variants are at our
disposal: i) geometrical collision criteria as employed in standard
hadronic transport, ii) the in-cell method developed in Ref.
\cite{Lang}. The latter can easily be extended to describe $2
\leftrightarrow 3$ processes etc. in a covariant way
\cite{NPA700}.
It is the better choice at high particle densities (cf. Ref.
\cite{XU}) and was actually used in the calculations presented below.
The hadronization is performed by integrating the rate equations
(e.g. (\ref{trans})) in space and time which are discretized on a
four-dimensional grid by $\Delta t$ and $\Delta V(t) = \Delta x(t)
\Delta y(t) \Delta z(t) $. In beam direction we use an initial grid
size $\Delta z = 1/\gamma_{cm}$ fm with $\gamma_{cm}$ denoting the
Lorentz-$\gamma$ factor in the nucleon-nucleon center-of-mass system
while in the transverse direction we use $\Delta x = \Delta y$ = 1
fm. The grid size is increased dynamically during the transport
calculation such that all particles are included on the actual grid.
This practically implies that the grid boundary in beam direction
approximately moves with the velocity of light. In each time step
$\Delta t$ and cell $\Delta V$ the integrals in (\ref{trans}) and
the respective integrals for baryon (antibaryon) formation
 are evaluated by a sum over all (time-like)
testparticles using (e.g. for the quark density)
\begin{equation} \label{rho_DV}
\phantom{a}\hspace*{-10mm}
\frac{1}{\Delta V} \int_{\Delta V} d^3x  \int_{-\infty}^\infty \frac{d
\omega_q}{2 \pi} \omega_q  \int_{-\infty}^\infty \frac{d^3  p_q}{(2\pi)^3} \
\rho_q(\omega_q,p_q)\ {\tilde N}_q(x,p_q) = \frac{1}{\Delta
V} \sum_{J_q \ {\rm in} \  \Delta V}  1 \ =  \ \rho_q(\Delta V) \ ,
\end{equation}
where the sum over $J_q$ implies a sum over all testparticles of
type $q$ (here quarks) in the local volume $\Delta V$ in each
parallel run. In Eq. (\ref{rho_DV}) ${\tilde N}$ denotes the occupation number
in phase space which in thermal equilibrium is given by Bose- or
Fermi-functions, respectively.
In case of other operators like the scalar density,
energy density etc. the number 1 in Eq. (\ref{rho_DV}) has to be
replaced by $\sqrt{P^2_J}/\omega_J$, $\omega_J$ etc.  In order to
obtain lower numerical fluctuations the integrals are averaged over
the  parallel runs (typically  50 at RHIC energies).
For each individual testparticle (i.e. $x_q$ and $p_q$ fixed) the
additional integrations in (\ref{trans})  give a probability for a
hadronization process to happen; the actual event then is selected
by Monte Carlo. Energy-momentum conservation fixes the
four-momentum $p$ of the hadron produced and its space-time position
$x$ is determined by (\ref{trans}). The final state is either a hadron
with flavor content fixed by the fusing quarks
(and/or antiquarks) or by a string of invariant mass $\sqrt{s}$ (with the same flavor), if
$\sqrt{s}$ is above 1.3 GeV for mesonic or above 1.5 GeV for baryonic quark content.

On the partonic side the following elastic and inelastic
interactions are included in PHSD $qq \leftrightarrow qq$, $\bar{q} \bar{q}
\leftrightarrow \bar{q}\bar{q}$, $gg \leftrightarrow gg$, $gg
\leftrightarrow g$, $q\bar{q} \leftrightarrow g$  exploiting
'detailed-balance' with interaction rates again from the DQPM
\cite{Cassing07,NPA09}. Numerical tests of the parton dynamics with respect to
conservation laws, interaction rates in and out-off equilibrium in a finite box
with periodic boundary conditions have been presented in Ref.
\cite{Vitaly}.
For further details we refer the reader to Section 2.2 of
\cite{NPA09}. The interactions between hadrons are the same as in
the HSD transport model.

\subsection{Initial conditions}
The initial conditions for the parton/hadron dynamical system have
to be specified additionally.  In order to describe relativistic
heavy-ion reactions we start with two nuclei in their
'semi-classical' groundstate, boosted towards each other with a
velocity $\beta$ (in $z$-direction), fixed by the bombarding energy.
The initial phase-space distributions of the projectile and target
nuclei are determined in the local Thomas-Fermi limit as in the HSD
transport approach \cite{PR90,Ehehalt,HSD} or the UrQMD model \cite{UrQMD1,UrQMD2}.
We recall that at
relativistic energies the initial interactions of two nucleons are
well described by the excitation of two color-neutral strings which
decay in time to the known hadrons (mesons, baryons, antibaryons)
\cite{JETSET}. Initial hard processes - i.e. the short-range high-momentum
transfer reactions that can be well described by perturbative QCD -
are treated in PHSD (as in HSD) via PYTHIA
5.7 \cite{PYTHIA}.
The novel element in PHSD (relative to HSD) is the 'string melting
concept' as also used in the AMPT model \cite{AMPT} in a similar
context. However, in PHSD the strings (or possibly formed hadrons)
are only allowed to 'melt' if the local energy density $\epsilon(x)$
(in the local rest frame) is above  the transition energy density
$\epsilon_c$. The present DQPM version (fitted to the lQCD results
from Ref. \cite{aori10}) gives $\epsilon_c \approx 0.5 $
GeV/fm$^3$. The mesonic strings then decay to quark-antiquark pairs
according to an intrinsic momentum distribution, \begin{equation}
\label{mom0} F({\bf q}) \sim \exp(- 2 b^2 {\bf q}^2) \ ,
\end{equation} in the meson rest-frame (cf. Eq. (\ref{trans}) for
the inverse process). The parton final four-momenta are selected
randomly according to the momentum distribution (\ref{mom0}) (with
$b$= 0.66 fm), and the parton-energy distribution is fixed by the
DQPM at given energy density $\epsilon(\rho_s)$ in the local cell with
scalar parton density $\rho_s$. The flavor content of the $q\bar{q}$
pair is fully determined by the flavor content of the initial
string. By construction the 'string melting' to massive partons
conserves energy and momentum as well as the flavor content. In
contrast to Ref. \cite{AMPT} the partons are of finite mass - in
line with their local spectral function - and obtain a random color
$c= (1,2,3)$ or $(r,b,g)$ in addition. Of course, the color
appointment is color neutral, i.e. when selecting a color $c$ for
the quark randomly the color for the antiquark is fixed by $-c$.
The baryonic strings melt analogously into a quark and a diquark
while the diquark, furthermore, decays to two quarks.

\subsection{Shadowing}
As well known from deeply-inelastic lepton scattering on nuclei,
there is a depletion of low-momentum partons in a nucleon embedded
in a nucleus compared to the population in a free nucleon, which
leads to a lowering in the production cross section of hard probes
in proton-nucleus and nucleus-nucleus collisions at high bombarding
energies compared to the production in the superposition of
independent nucleon-nucleon collisions. The reasons for depletion,
though, are numerous, and models of shadowing vary accordingly.
There is, therefore, a considerable (about a factor of 3)
uncertainty in the amount of shadowing predicted at RHIC and
especially at LHC
\cite{Gyul91,Vogt,CapellaGluon,Bravina,Kopeliovich,Kari1,Kari2,Kari3}.
In the analysis of the $d+Au$ data at $\sqrt{s_{NN}}=200$~GeV, in
which the maximum estimate for the effect of the shadowing was
made~\cite{Vogt,Kari3,dA}, the shadowing lead to a $\sim$ 10\%
reduction while an anti-shadowing closer to target rapidities was
observed. More recent estimates for the shadowing at RHIC energies
have been presented in Refs. \cite{Kari1,Kari2} and imply a lower
amount of shadowing for hard probes. In the present PHSD
calculations we employ the results of Ref. \cite{Gyul91} which lead
to a suppression of charmonia at forward rapidities in d+Au
reactions due to shadowing by less than 5\%.

The actual implementation of shadowing in PHSD is done in a practical
way by parametrizing the suppression (or enhancement) factors from Ref.
\cite{Gyul91} for heavy nuclei at RHIC energies as a function of the
Bjorken variable $x \sim 2 p_0/\sqrt{s}$ and the mass number $A$, i.e.
\begin{eqnarray}
\label{shadow}
R_A(x) &=& 1 + 1.19 \ln^{1/6} [x^3 -
1.5 (x_0+x_L) x^2 + 3 x_0 x_L x]    \\
       &-& \left[ \alpha _A -
\frac{1.08 (A^{1/3}-1)}{\ln(A+1)} \sqrt{x}
\right] \exp(-x^2/x_0^2) ,
\nonumber\end{eqnarray}

with the parameters $x_0 = 0.1, x_L = 0.7$ and $\alpha_A = 0.1
(A^{1/3}-1)$. Since Eq. (\ref{shadow}) determines only the average
nuclear effect one has to specify the impact parameter dependence
of the shadowing function $R_A(x)$ on the actual position of the
colliding nucleons. As in Ref. \cite{Gyul91} we assume that the
shadowing parameter $\alpha_A$ is proportional to the longitudinal
thickness of the nucleus at impact parameter $r$ and adopt
\begin{equation} \label{shadow2} \alpha_A(r) = 0.1 (A^{1/3} -1)
\ \frac{4}{3} \ \sqrt{1-r^2/R_A^2} ,
\end{equation} where $r$ is the transverse distance of the
interacting nucleon from its nucleus center and $R_A$ is the
radius of its nucleus. In this manner an
approximate implementation of initial state shadowing is achieved which
is sufficient for the purposes of the present investigations.  Since the
shadowing effects are only on the level of a few percent for the
observables addressed here we discard a further discussion and shift
its representation to a forthcoming study of hard probes.

Let's summarize the modifications and extensions of the PHSD approach
relative to the version presented before: except
for shadowing and the new partonic
equation-of-state from Ref. \cite{aori10}, we adopt here essentially the same
approach as described in detail in Ref. \cite{NPA09}, which we use now to investigate
heavy-ion reactions at the top RHIC energy while in Ref. \cite{NPA09} the SPS energy
regime was studied.

\section{Application to Au + Au collisions at $\sqrt{s}$ = 200 GeV}
In this Section we employ the PHSD approach - described briefly in Section 2
and in more detail in Ref. \cite{NPA09} - to nucleus-nucleus collisions at
ultra-relativistic energies, i.e. in particular at the top RHIC energy. Note that at
RHIC or more specifically LHC energies other initial conditions
(e.g. a color-glass condensate \cite{Larry}) might be necessary. In
the present work we discard such alternative initial conditions and
explore to what extent the present initial conditions are compatible
with differential measurements by the various collaborations at RHIC.

\subsection{Parton dynamics at RHIC energies}
We start with a consideration of energy partitions in order to map
out the fraction of partonic energy in time for relativistic
nucleus-nucleus collisions. In Ref. \cite{NPA09} we have found that
even in central collisions of Pb +Pb at 158 A GeV only a limited
fraction of degrees-of-freedom can be attributed to a partonic phase
due to a significant hadronic corona \cite{Werner,Werner2} both in
coordinate space as well as for large rapidities. In order to address
the experimental observations at RHIC we will focus here on the
total energy at midrapidity, i.e. for $|y| \leq 1$. We note in
passing that the total energy - integrated over all rapidities - is
conserved throughout the reaction within better than 1\%
(cf. also Ref. \cite{NPA09}).

\begin{figure}[tb]
\centering \includegraphics*[width=95mm]{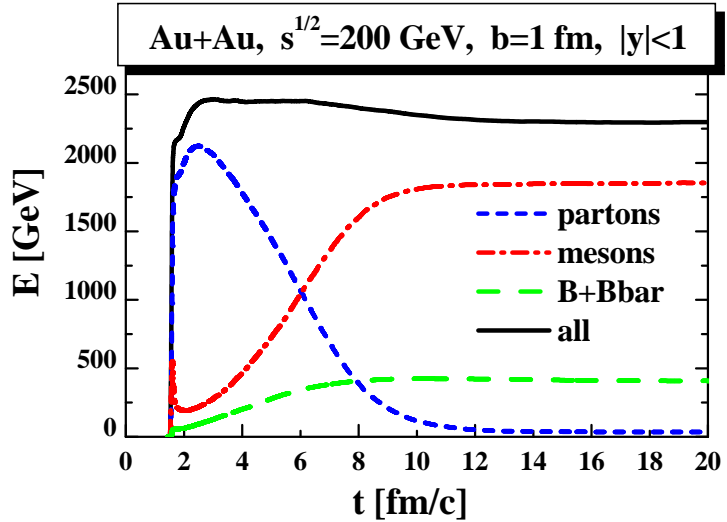} \caption{The total energy $E_{tot}$
(upper solid black line) for central ($b$=1 fm) collisions of Au+Au at $\sqrt{s}$ = 200 GeV
within the rapidity window $|y| \leq 1$.
The dashed (blue) line shows the energy contribution from
partons while the dot-dashed (red) line displays the energy
contribution from mesons (including 'unformed mesons' in strings).
The long-dashed (green) line is the contribution of
baryons (and antibaryons).  } \label{fig8}
\end{figure}

In Fig. \ref{fig8} we show the energy balance for a central (impact
parameter $b$=1 fm) reaction of Au+Au at $\sqrt{s}$ = 200 GeV, i.e.
at the top RHIC energy including partonic, mesonic and baryonic
degrees of freedom in the rapidity window $|y| \leq 1$. The total
energy $E_{tot}$ within this rapidity interval (upper line) - which
at $t=0$ is zero in the cms of the colliding nuclei - shows a rapid
increase in time for $t \approx $ 1.6 fm/c which corresponds to the
contact time of the colliding heavy ions. At $t \approx $ 2.5 fm/c
about 85\% of the energy (at midrapidity) is carried by the partonic
degrees-of-freedom which are converted with increasing time to
mesons and baryons (or antibaryons) essentially within 6-8 fm/c. The
total energy within this rapidity window is not conserved since by
elastic and inelastic reactions the reaction products may leave (or
enter) the rapidity window. Note that even in central collisions (at
midrapidity) not all the energy is converted to a partonic phase and
a hadronic (or rather string-like) corona \cite{Werner,Werner2}
survives in the surface area of the collision zone.

We note in passing that a qualitatively similar picture is
obtained when plotting the parton, meson and baryon (+ antibaryon)
numbers for $|y| \leq 1$.  An essential
point here is that the number of final hadronic states is larger
than the maximal number of partons, i.e. there is a production of entropy in
the hadronization process as pointed out before in Refs.
\cite{PRC08,NPA09} and thus no violation of the second law of thermodynamics in PHSD!

\begin{figure}[tb]
\centering \includegraphics*[width=85mm]{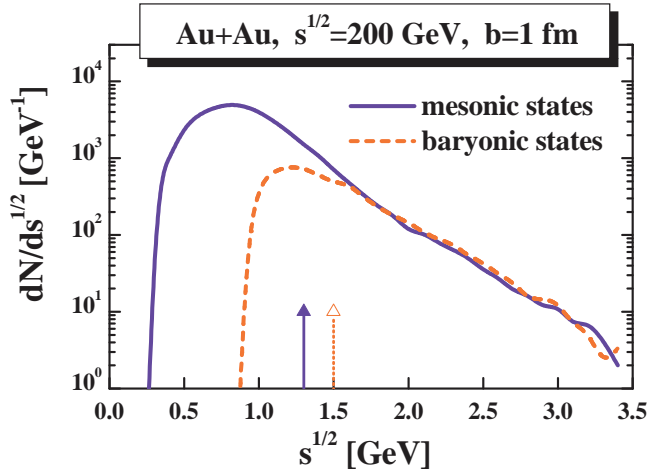} \caption{The
invariant mass distribution for fusing $q \bar{q}$ pairs (solid
blue line) as well as $qqq$ (and $\bar{q}\bar{q}\bar{q}$) triples
(dashed red line) that lead to the formation of final hadronic states
for Au+Au at $\sqrt{s}$ = 200 GeV ($b$=1 fm). The vertical arrows denote the
thresholds for mesonic (1.3 GeV) and baryonic strings (1.5 GeV). }
\label{fig12a}
\end{figure}

In order to shed more light on the hadronization process in
PHSD we display in Fig. \ref{fig12a} the invariant mass
distribution of $q \bar{q}$ pairs (solid line) as well as $qqq$
(and $\bar{q}\bar{q}\bar{q}$) triples (dashed line) that lead to
the formation of final hadronic states. The reaction is again
Au + Au at $\sqrt{s}$ = 200 GeV at impact parameter $b$= 1 fm.
In fact, the distribution for the
formation of baryon (antibaryon) states starts above the nucleon
mass and extends to high invariant mass covering the nucleon
resonance mass region as well as the high-mass continuum (which is
treated by the decay of strings within the JETSET model
\cite{JETSET,Falter} for $M >$ 1.5 GeV). On the 'pre-mesonic' side the invariant-mass
distribution starts  above the pion mass and extends up
to continuum states of high invariant mass (described again in
terms of string excitations for $M > $ 1.3 GeV). The low-mass sector is dominated by
$\sigma, \rho$, $a_1$, $\omega$ or $K^*, \bar{K}^*$ transitions etc. As
mentioned before the excited 'pre-hadronic' states  decay to two
or more 'pseudoscalar octet' mesons such that the number of final
hadrons is larger than the initial number of fusing partons.

\begin{figure}[tb]
\centering \includegraphics*[width=85mm]{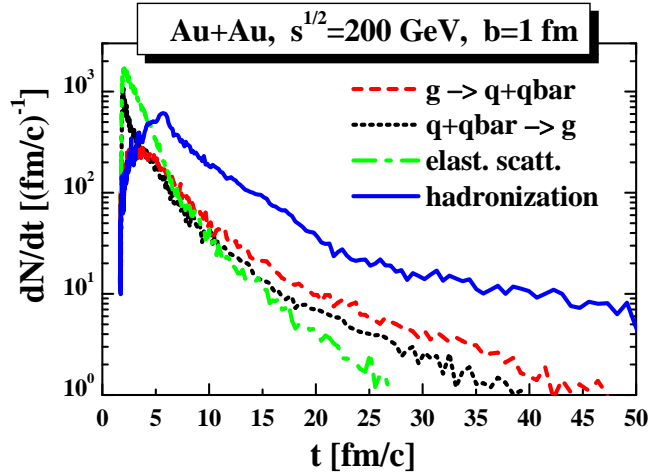} \caption{The
interaction rates (integrated over rapidity) for the channels $ q
\bar{q}  \rightarrow g$ (black dot-dashed line), $ g \rightarrow q
\bar{q} $ (dashed red line) and elastic parton scattering (dashed
green line). The solid blue line displays the hadronization rate.
The reaction is Au+Au at $\sqrt{s}$ = 200 GeV ($b$=1 fm). The
rateinclude all rapidities.} \label{fig13a}
\end{figure}

The individual reaction rates in the partonic phase are of further
interest in central Au+Au collisions at the top RHIC energy. To this
aim we display in Fig. \ref{fig13a} the interaction rates for the
channels $ q \bar{q} \rightarrow g$ (black dotted line), $g
\rightarrow  q \bar{q} $ (dashed red line) and elastic parton
scattering (dot-dashed green line). Except for the very early phase,
where the elastic scattering channels dominate, all interaction
rates are of comparable size and decrease rapidly in time. For
comparison we also show the hadronization rate by the solid blue
line which has a maximum at about 6 fm/c and is sizeably larger than
the other interaction rates for $t > $ 7 fm/c. Note that in this
representation we have considered all rapidities; accordingly
interaction and hadronization processes keep on going at forward and
backward rapidities also for times larger than 50 fm/c.

\begin{figure}[tbh]
 \includegraphics*[width=75mm]{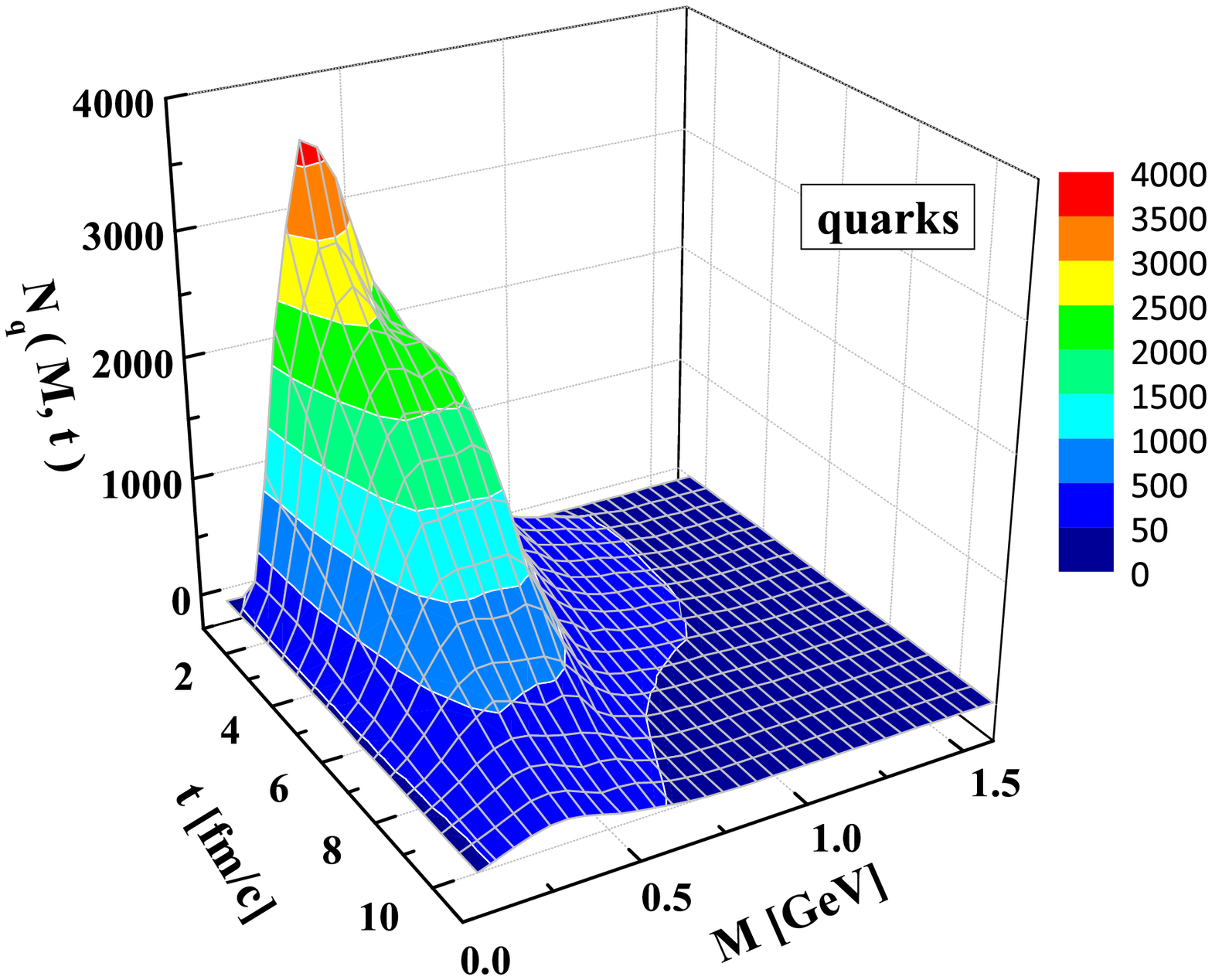} \includegraphics*[width=75mm]{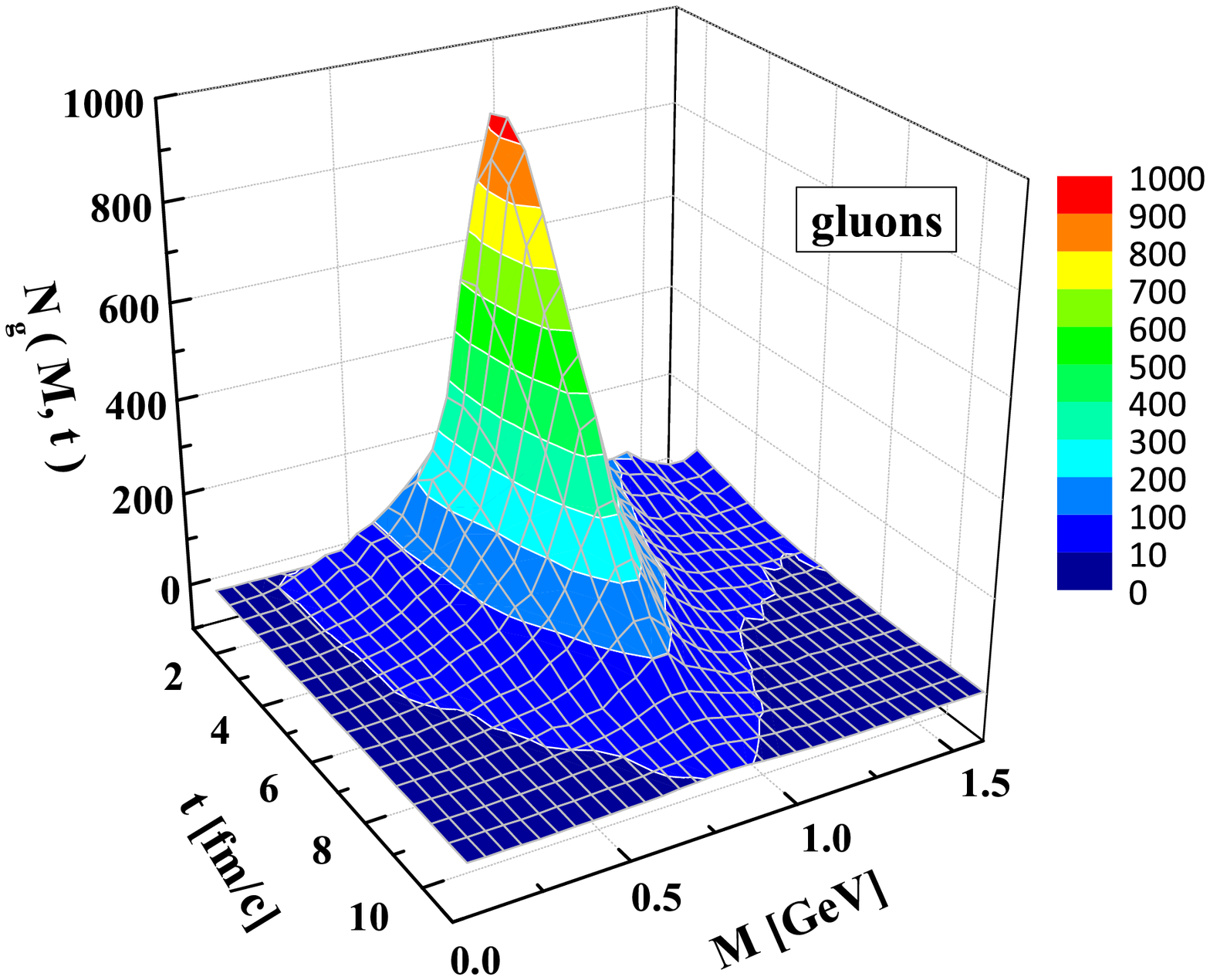}
 \caption{The
time-dependent mass distributions for quarks (+ antiquarks)
(l.h.s.) and gluons (r.h.s.) for a central Au+Au collision at
$\sqrt{s}$ = 200 GeV and $b$=1 fm at midrapidity ($|y| \leq 1$). }
\label{speq}
\end{figure}

Some information on the time evolution of the quark and gluon mass
functions is displayed in Fig. \ref{speq} which shows the number of
'particles' as a function of invariant mass $M$ and time $t$ at
midrapidity ($|y| \leq 1$). Note that by integration over $M$ one
obtains the number of quarks (+ antiquarks) $N_q(t)$ and gluons
$N_g(t)$ in the rapidity interval $|y| \leq 1$ while dividing by
$N_q(t)$ and $N_g(t)$, respectively, a rough estimate for the
particle spectral functions is obtained. Note that the mass
distributions displayed here are the product of the spectral
functions and the occupation numbers in a restricted phase space.
Due to a moderate variation of the partons pole mass and width with
the scalar density $\rho_s$ (cf. Fig. 2) the shapes of the partonic
mass distributions do not change very much in time. The average
quark mass is about 0.5 GeV while the average gluon mass is only
slightly less than 1 GeV. Note, however, that the width of the mass
function - which reflects the actual interaction rate per parton -
 remains significant for all times up to hadronization. Furthermore, the
average parton width as a function of time cannot directly be
related to Figs. 1 or 2 since at a given time $t$ the partonic
mass distribution relates to different scalar densities in the
course of the partonic evolution.

\subsection{Particle spectra in comparison to experiment}

\begin{figure}[tbh]
\centerline{\psfig{figure=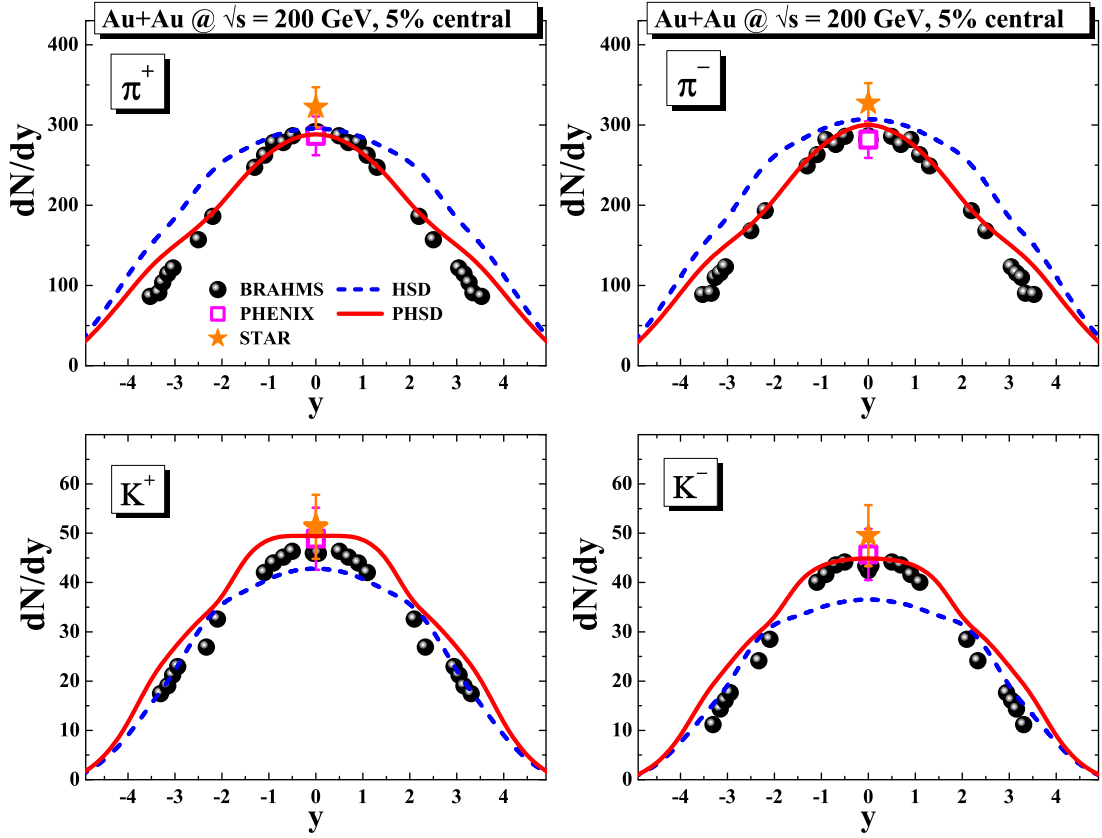,width=15.cm}} \caption{The
rapidity distribution of $\pi^+$ (upper part, l.h.s.), $K^+$
(lower part, l.h.s.), $\pi^-$ (upper part, r.h.s.) and $K^-$
(lower part, r.h.s.)  for  5\% central Au+Au collisions at
$\sqrt{s}$ = 200 GeV from PHSD (solid red lines) in comparison to
the distribution from HSD (dashed blue lines) and the experimental
data from the RHIC Collaborations \cite{PHENIX2,STAR3,BRAHMS}. } \label{fig13}
\end{figure}

Apart from the more general considerations in the previous
Subsection, it is of interest how the PHSD approach compares to
the HSD model (without explicit interacting partonic degrees of freedom) as
well as to experimental data from the RHIC collaborations. We
start with rapidity spectra from PHSD (solid red lines) for
charged pions and kaons in 5\% central Au+Au collisions at
$\sqrt{s}$ = 200 GeV which are compared in Fig. \ref{fig13} to the
data from the RHIC Collaborations \cite{PHENIX2,STAR3,BRAHMS} as well as
to results from
HSD (dashed blue lines). We find the rapidity distributions of the
charged mesons to be slightly narrower than those from HSD and
actually closer to the experimental data. Also note that there is
slightly more production of $K^\pm$ mesons in PHSD than in HSD
while the number of charged pions is slightly lower. The actual
deviations between the PHSD and HSD spectra are not dramatic but
more clearly visible than at SPS energies (cf. Ref. \cite{NPA09}).
Nevertheless, it becomes clear from Fig. \ref{fig13} that the
energy transfer in the nucleus-nucleus collision from initial nucleons
to produced hadrons - reflected dominantly in the light meson spectra
- is rather accurately described by PHSD. Fig. \ref{fig13} also
demonstrates that the longitudinal motion is well understood
within the PHSD approach.

\begin{figure}[tbh]
\centerline{\psfig{figure=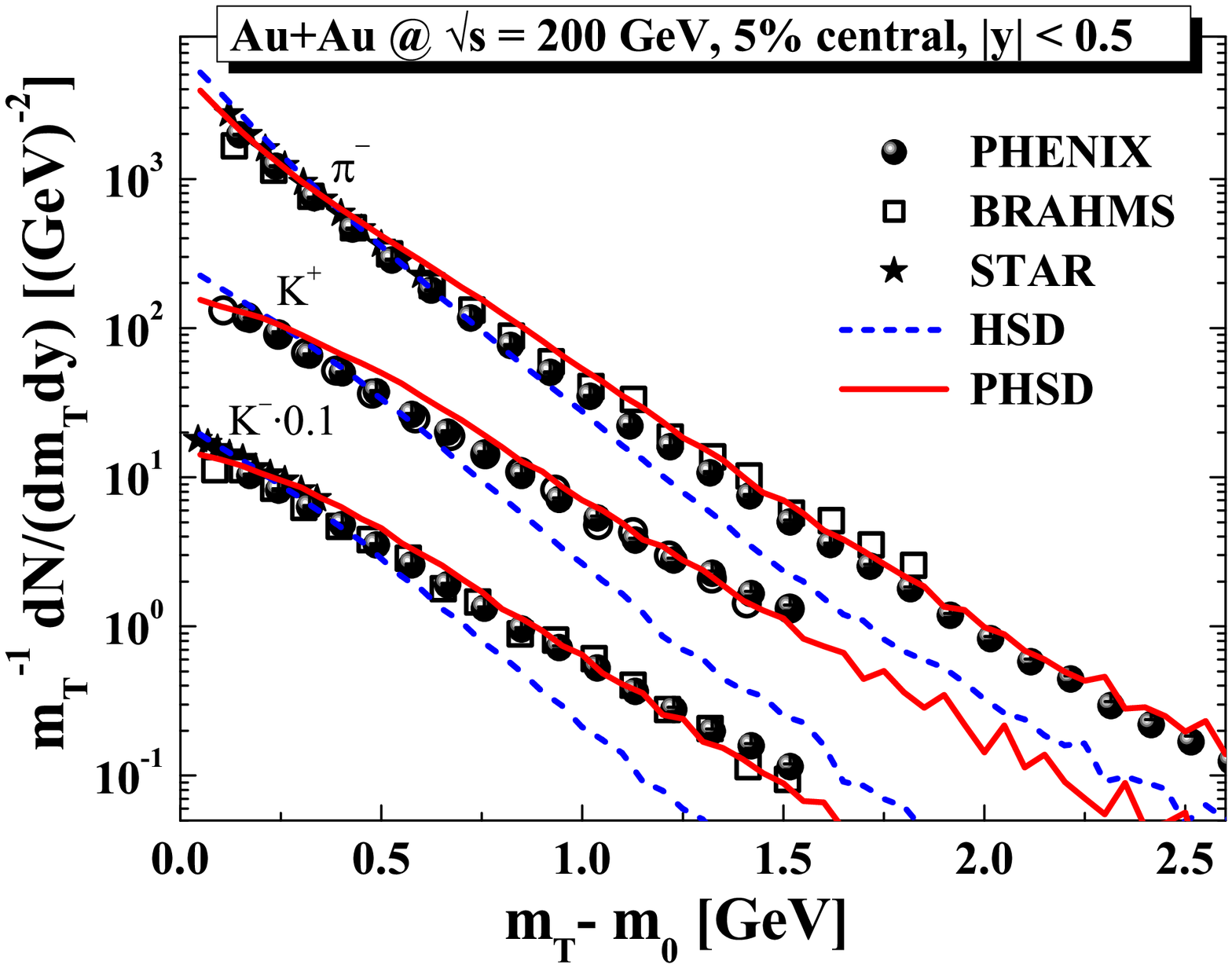,width=9.cm}} \caption{The
$\pi^-$, $K^+$ and $K^-$ transverse mass spectra for 5\% central
Au+Au collisions at $\sqrt{s}$ = 200 GeV from PHSD (solid red lines)
in comparison to the distributions from HSD (dashed blue lines) and
the experimental data from the BRAHMS, PHENIX and STAR
Collaborations \cite{PHENIX2,STAR3,BRAHMS} at midrapidity.  }
\label{fig14}
\end{figure}

Independent information on the active degrees of freedom is provided by
transverse mass spectra of the hadrons especially in central collisions.
The actual results for RHIC energies are displayed in Fig.  \ref{fig14} where we
show the transverse mass spectra of $\pi^-$, $K^+$  and $K^-$ mesons
for 5\% central Au+Au collisions at $\sqrt{s}$ = 200 GeV in comparison to the data of the
RHIC Collaborations
\cite{PHENIX2,STAR3,BRAHMS}.  Here the slope of the $\pi^-$ spectra
is slightly enhanced in PHSD (solid red lines) relative to HSD
(dashed blue lines)  which demonstrates that the pion transverse
mass spectra
also show some sensitivity to the partonic phase (contrary to the SPS energy regime). The
$K^\pm$ transverse mass spectra are substantially hardened with respect
to the HSD calculations - i.e. PHSD is more in line with
the data - and thus suggest that partonic effects are better visible in
the strangeness degrees-of-freedom. The hardening of the kaon spectra
can be traced back to parton-parton scattering as well as a larger
collective acceleration of the partons in the transverse direction due
to the presence of the repulsive mean-field for the partons (cf. Fig. 5).
The enhancement of the spectral slopes for kaons and antikaons in PHSD
due to collective partonic flow shows up much clearer for the kaons due
to their significantly larger mass (relative to pions).

\begin{figure}[tbh]
\centerline{\psfig{figure=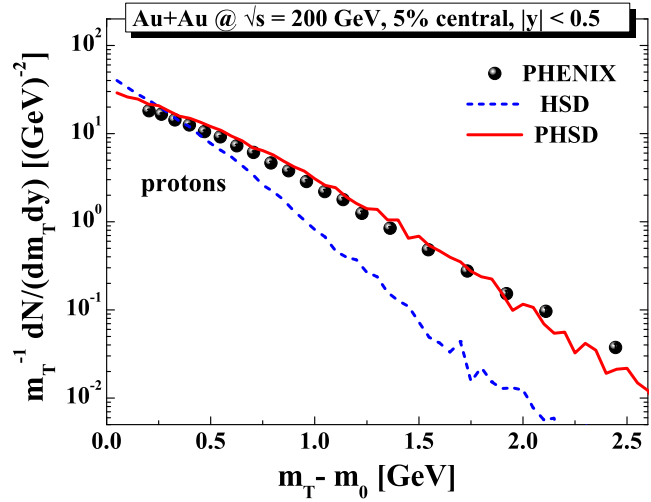,width=9.cm}} \caption{The
proton transverse mass spectra for 5\% central Au+Au collisions at
$\sqrt{s}$ = 200 GeV from PHSD (solid red line) in comparison to the
distributions from HSD (dashed blue line) and the experimental data
from the PHENIX Collaboration \cite{PHENIX2} at midrapidity.  }
\label{pspectra}
\end{figure}

The latter considerations also become transparent when comparing
the transverse mass spectra for protons at midrapidity from HSD
and PHSD to the data from the PHENIX Collaboration \cite{PHENIX2} in Fig.
\ref{pspectra}. Here the HSD spectra (dashed blue line) severely
underestimate the slope of the data from Ref. \cite{PHENIX2} whereas
the PHSD spectra (solid red line) are fairly in line.
These differences are so dramatic because in HSD the protons
at midrapidity dominantly stem from initial string decays and are
not allowed to rescatter during their formation time of $\gamma_L
\tau_0$ where $\gamma_L$ denotes the Lorentz factor and $\tau_0$ =
0.8 fm/c is the default formation time for hadrons in HSD
\cite{Ehehalt,HSD}. On the other hand, in PHSD the dominant source
of protons (and also the other baryons and antibaryons) at midrapidity is the
fusion of three quarks from the partonic phase. Since the partonic
degrees of freedom interact strongly and are accelerated in the
expansion phase due to the repulsive mean field the protons pick
up the momenta from the fusing partons and thus show a sizeable
harder slope in the transverse mass spectrum.

In summarizing this Subsection we point out that the partonic
phase in PHSD at the top RHIC energy leads to a narrowing of the
longitudinal momentum distribution, a reduction of pion production
and slight enhancement of kaon production  and to a hardening of
their transverse mass spectra relative to HSD (closer to the data).
These effects are  clearly visible especially in the transverse
degrees-of-freedom and are identical to those of
Ref. \cite{NPA09} at SPS energies, however, more pronounced due to
the larger space-time region of the partonic phase.

\subsection{Elliptic flow}
Of additional interest are the collective properties of the
strongly interacting system which are explored experimentally via
the elliptic flow
\begin{equation} v_2(p_T,y) = \left\langle (p_x^2 - p_y^2)/(p_x^2 +
p_y^2) \right\rangle |_{p_T,y}
\end{equation} of hadrons as a function of centrality, rapidity $y$, transverse
momentum $p_T$
or transverse kinetic energy per participating quarks and
antiquarks. We note that the reaction plane in PHSD is given by the $x-z$ plane
with the $z$-axis in beam direction.

\begin{figure}[tbh]
\centering \includegraphics*[width=105mm]{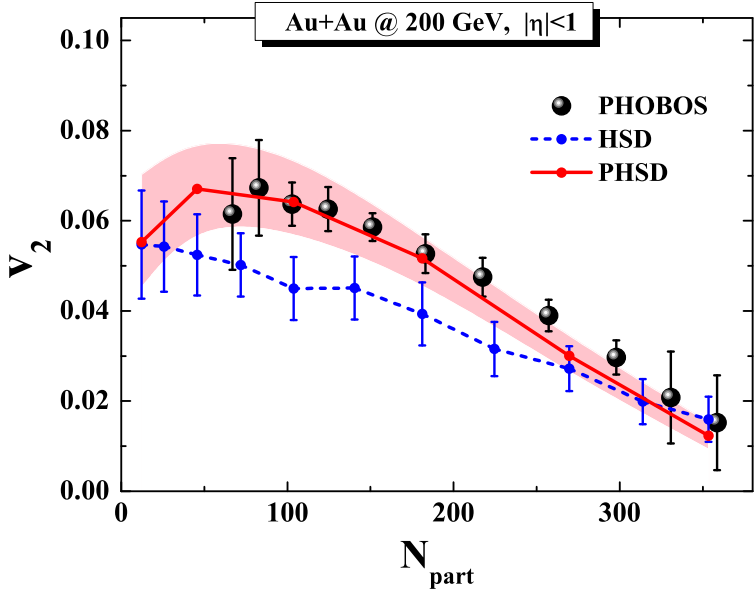}
\caption{ The elliptic flow $v_2$ for Au+Au collisions at the top
RHIC energy $\sqrt{s}$ = 200 GeV  as a function of the centrality
 measured by the number of participating nucleons
$N_{part}$. The solid (red) line stands for the results from PHSD
whereas the dashed (blue) line represents the results from HSD (from Ref. \cite{Brat03}).
The data are taken from the PHOBOS Collaboration \cite{PHOBOS4} and
correspond to momentum integrated events in the pseudo-rapidity window
$|\eta| \leq 1$ for charged particles. The shaded band signals the statistical uncertainties of the
PHSD calculations.}
 \label{fig22}
\end{figure}

We start in Fig. \ref{fig22} with the elliptic flow $v_2$ (for Au+Au
collisions at the top RHIC energy) as a function of the centrality
of the reaction measured by the number of participating nucleons
$N_{part}$. The solid (red) line stands for the results from PHSD
which is compared to the data for charged particles from the PHOBOS
Collaboration \cite{PHOBOS4}. The dashed blue line refers to the
corresponding results for $v_2$ from HSD (taken from Ref.
\cite{Brat03}). The momentum integrated results in the
pseudo-rapidity window $|\eta| \leq 1$ from PHSD compare well to the
data from Ref. \cite{PHOBOS4} whereas the HSD results clearly
underestimate the elliptic flow as pointed out before in Ref.
\cite{Brat03}. The relative enhancement of $v_2$ in PHSD with
respect to HSD can be traced back to the high interaction rate in
the partonic phase and to the repulsive mean field for partons (cf.
Fig. 5). We note in passing that PHSD calculations without mean
fields only give a tiny enhancement for the elliptic flow relative
to HSD.

\begin{figure}[tbh]
\centering \includegraphics*[width=105mm]{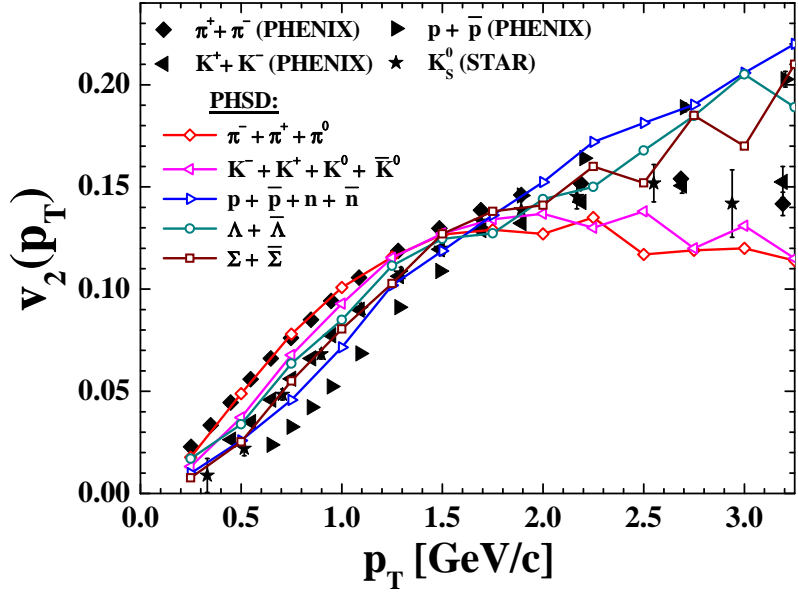}
\caption{ The hadron elliptic flow $v_2$ for
inclusive Au+Au collisions as a function of the transverse
momentum $p_T$ (in GeV) for different hadrons in comparison to the data from the STAR \cite{STAR6}
and PHENIX Collaborations \cite{SCALING2} within the same rapidity cuts. } \label{fig23}
\end{figure}

Fig. \ref{fig23} shows the final hadron $v_2$ versus the
transverse momentum $p_T$ for different particle species in
comparison to the data from the STAR \cite{STAR6}
and PHENIX Collaborations \cite{SCALING2}. We observe a mass
separation in $p_T$ as well as a separation in mesons and baryons
for $p_T >$ 2 GeV roughly in line with data. The elliptic flow of
mesons is slightly underestimated for $p_T >$ 2 GeV in PHSD which is
opposite to ideal hydrodynamics which overestimates $v_2$ at high
transverse momenta. On the other hand, the proton (and antiproton)
elliptic flow is slightly overestimated at low transverse momenta $p_T < $ 1.5 GeV.

\begin{figure}[tbh]
\centering \includegraphics*[width=105mm]{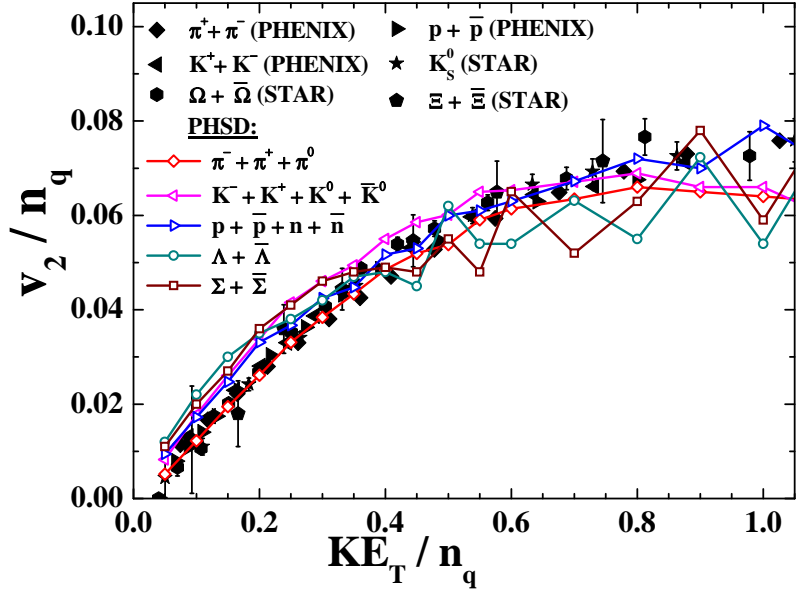}
\caption{The elliptic flow $v_2$ - scaled by the number of
constituent quarks $n_q$ - versus the transverse kinetic energy
(\ref{ETR}) divided by $n_q$ for different hadron species in comparison to the data from the STAR \cite{STAR6}
and PHENIX Collaborations \cite{SCALING2}. } \label{fig25}
\end{figure}

A further test of the PHSD hadronization approach is provided by
the  'constituent quark number scaling' of the elliptic flow $v_2$
which  has been observed experimentally in central Au+Au
collisions at RHIC \cite{STARS,SCALING2,SCALING1}. In this respect
we plot $v_2/n_q$ versus the transverse kinetic energy per
constituent parton,
\begin{equation} \label{ETR} KE_T = \frac{m_T-m}{n_q} \ , \end{equation}
with $m_T$ and $m$  denoting the transverse mass and actual hadron
mass, respectively. For mesons we have $n_q = 2$ and for
baryons/antibaryons $n_q=3$. The results for the scaled elliptic
flow are shown in Fig. \ref{fig25} in comparison to the data from
the STAR \cite{STAR6} and PHENIX Collaborations \cite{SCALING2} for
different hadrons and suggest an approximate scaling. For $KE_T >
0.5$ GeV there is a tendency to underestimate the experimental
measurements for $\Lambda, \Sigma, \bar{\Lambda}, \bar{\Sigma}$
baryons which we attribute to an underestimation of interaction
terms in PHSD for high momentum hadrons. In this respect we recall
that the momentum independence of the quasiparticle width $\gamma$
and mass $M$ (cf. Subsection 2.1) is presently a rough approximation
and has to be refined.   Due to the limited statistics especially in
the baryonic sector with increasing $p_T$ this issue will have to be
re-addressed with high statistics in future.

 \vspace{1cm}

\section{Summary and outlook}

In summary, relativistic collisions of Au+Au at top RHIC energies
have been studied within the PHSD approach which includes explicit
partonic degrees of freedom as well as dynamical local transition
rates from partons to hadrons (cf. Eq. (\ref{trans})). The partonic
equation-of-state employed has been adopted from the lQCD
calculations of the Wuppertal/Budapest group in thermodynamic
equilibrium \cite{aori10} and incorporated in the PHSD approach by
means of the Dynamical QuasiParticle Model (DQPM)
\cite{Cassing07}. The transition from partonic to hadronic degrees
of freedom is described by covariant transition rates for the
fusion of quark-antiquark pairs or three quarks (antiquarks),
respectively, obeying flavor current-conservation, color
neutrality as well as energy-momentum conservation.

Our dynamical studies for heavy-ion collisions at relativistic
collider energies have been compared to earlier results from the
Hadron-String Dynamics (HSD) approach - incorporating no explicit
interacting partonic phase - as well as to experimental data from
the STAR, PHENIX, BRAHMS and PHOBOS collaborations for Au+Au
collisions at the top RHIC energy of $\sqrt{s}$ = 200 GeV. We find a
reasonable reproduction of hadron rapidity distributions and
transverse mass spectra and also an acceptable description of the
elliptic flow of charged hadrons as a function of the centrality of
the reaction and the transverse momentum $p_T$. This result is quite
remarkable since the additional interactions of partonic nature
(relative to HSD) are essentially determined by the DQPM which
itself has been fixed by lQCD 'data'.  Furthermore, an approximate
'quark-number scaling' of the elliptic flow $v_2$ is observed in the
PHSD results, too, while the HSD calculations underestimate the
elliptic flow observables. We mention that no fitting  has been
addressed in our study and that the discrepancies with respect to
the data of the RHIC collaborations might be attributed to the crude
approximations with respect to the quasiparticle mass and width,
i.e. in particular by adopting momentum-independent quantities. This
definitely needs improvement and more detailed investigations in
future.

Since the bulk dynamics of nucleus-nucleus reactions at RHIC
energies appear to be reasonably  described by the PHSD
approach future studies will concentrate on leptonic and photonic
probes as well as charm and high $p_T$ degrees of freedom where
previous studies within HSD showed sizeable discrepancies with
respect to the data taken at the SPS or particularly at RHIC
energies \cite{Lena09,charmrev,Voda10} and the necessity for
partonic degrees of freedom had been pointed out. In fact, the
various leading channels in the partonic phase for dilepton
production have already been calculated on the basis of the DQPM
propagators in Ref. \cite{Linnyk11} and been implemented in PHSD.
Preliminary results have been presented in Ref. \cite{Olena11b} and
appear encouraging.

Furthermore, an application of PHSD at LHC energies will be
mandatory in order to map out the parton dynamics at even higher
energy densities. Since the DQPM predicts rather moderate changes
of the quasiparticle properties when increasing the temperature
from $\sim 2 T_c$ to about $4 T_c$ only minor changes in the
collective properties are expected within PHSD when comparing
nucleus-nucleus collisions at LHC relative to RHIC energies.
However, this expectation will have to be studied in detail in
comparison to the data that have become available recently
\cite{LHC1}.

\section*{Acknowledgement}
The authors are grateful to
R. Bellwied, L. Csernai, M. Gorenstein,  B. Jacak, R. Lacey,
C. Markert, V. Ozvenchuk, V.D. Toneev, V. Voronyuk and N. Xu
for valuable discussions. This work in part has been supported by
DFG as well as by the LOEWE center HIC for FAIR.


\end{document}